\documentclass[fleqn,usenatbib]{mnras}

\usepackage[T1]{fontenc}
\usepackage{amsmath}
\usepackage{amssymb}
\usepackage{bm}

\usepackage{graphicx}
\usepackage{acronym}
\usepackage{xcolor}
\definecolor{grahamcolour}{rgb}{0.64,0.20,0.00}
\usepackage{booktabs}
\hypersetup{colorlinks=true, allcolors=blue}
\usepackage[normalem]{ulem}

\acrodef{GW}[GW]{gravitational wave}
\acrodef{FFT}[FFT]{fast Fourier transform}
\acrodef{SNR}[SNR]{signal-to-noise ratio}
\acrodef{PSD}[PSD]{power spectral density}

\acrodef{ASD}[ASD]{amplitude spectral density}
\acrodef{OMS}[OMS]{optical metrology system}
\acrodef{LISA}[LISA]{Laser Interferometer Space Antenna}
\acrodef{GP}[GP]{Gaussian process}
\acrodef{RMS}[RMS]{root mean square}
\acrodef{SVD}[SVD]{singular value decomposition}
\acrodef{MOSA}[MOSA]{moving optical sub assembly}
\acrodef{KS}[KS]{Kolmogorov--Smirnov}

\begin{document}
\label{firstpage}
\pagerange{\pageref{firstpage}--\pageref{lastpage}}

\title{TDI-\(\infty\) and beyond}

\author[J. Bayley et al.]{
    Joe Bayley,$^{1}$\thanks{E-mail: joseph.bayley@glasgow.ac.uk}
    Graham Woan$^{1}$
    and Ollie Burke$^{1}$
    \\
    $^{1}$SUPA, University of Glasgow, Glasgow G12 8QQ, United Kingdom
}

\maketitle

\begin{abstract}

    Time-delay interferometry (TDI) cancels the laser frequency noise that dominates the measurements of the Laser Interferometer Space Antenna (LISA), and relies on interpolating the laser noise to fractional delays.
    In TDI-\(\infty\), which finds the laser noise free combinations from a linear model of the measurements, the interpolation sets the design matrix, and any interpolation error leaves uncanceled laser noise which is not accounted for.
    We introduce Gaussian process (GP) interpolation into TDI-\(\infty\), which models the laser noise from its spectrum and propagates its interpolation error into the likelihood as a covariance.
    We test it on a realistic data model in which the measurements pass through the on-board anti-alias filter and are downsampled to the 4 Hz telemetry rate, either applying the filter after the delays, as on the spacecraft, or building the filter into the GP kernel.
    We compare it with Lagrange and cosine-sum interpolation, and also keep the column space of the design matrix, which the standard null space likelihood discards.
    In \texttt{lisainstrument} simulations all three interpolators cancel the laser noise for a 1 Hz filter band edge, but only the GP does so up to the 2 Hz Nyquist frequency, a range the Lagrange and cosine-sum interpolators were not designed for.
    In parameter estimation on 200 simulated galactic binaries, the GP with the filter in its kernel gives the smallest calibrated posteriors.
    The differences are small for a single long data segment and grow when the data are split into short chunks or contain gaps, where the column space also recovers some information, by an amount that depends on the threshold that defines the null space.

\end{abstract}

\begin{keywords}
    gravitational waves -- methods: data analysis -- instrumentation: interferometers
\end{keywords}

\section{Introduction}

The \ac{LISA} \citep{amaro-seoane_laser_2017} is a future space-based gravitational wave detector that will observe gravitational waves in the millihertz band \citep{colpi_lisa_2024}.
It consists of three spacecraft forming a triangle of side \(\sim 2.5\) million kilometres,  connected by laser links.
Unlike ground-based detectors, the arm lengths are unequal and vary with time, so the frequency noise of the lasers does not cancel naturally and is orders of magnitude larger than any other noise source.
This laser noise must be removed before gravitational waves can be detected.

The noise from each laser is recorded in more than one measurement, since the beam from a given laser is beaten against a local laser both at its own spacecraft and, after propagating along an arm, at the distant spacecraft.
The same noise therefore appears in several measurements at different delays, and can be cancelled in post-processing by combining those measurements with the appropriate delays in a process called time-delay interferometry (TDI) \citep{tinto_time-delay_2021}.
Traditional TDI constructs fixed analytic combinations of the delayed measurements that cancel the laser noise \citep{armstrong_time-delay_1999}, and second-generation combinations extend this to the slowly time-varying \ac{LISA} arms \citep{tinto_second-generation_2023}.
Another approach, TDI-\(\infty\) \citep{vallisneri_tdi-infinity_2021}, builds on the principal-component view of laser-noise cancellation \citep{romano_principal_2006}. It writes the likelihood directly in terms of the raw measurements and marginalises over the laser noise in a linear model of the measurements, rather than using a fixed set of analytically constructed combinations.
This handles time-varying delays, data gaps, and additional primary noises such as optical-bench and clock noise within a single framework, where traditional TDI treats each with a separate processing step.

TDI-\(\infty\) has so far been demonstrated on simplified datasets, for example for tilt-to-length noise estimation \citep{houba_time-delay_2023} and for parameter inference in the presence of data gaps \citep{houba_robust_2024}.
In both of these, the on-board filtering and downsampling of the real instrument was not included, and the design matrix was built and applied at the same sample rate as the simulated data.
In a realistic \ac{LISA} setting the frequency or phase measurements are first made on board at a rate of 80\,MHz, then filtered by an anti-aliasing filter and downsampled to the telemetry rate of 4\,Hz before being sent to the ground \citep{bayle_unified_2023}. On board this decimation is done as a cascade of stages \citep{staab_laser_2023}. We model it as a single anti-alias filter, which gives the same effective band limit. In this work we reproduce this structure at a reduced physics rate, which is computationally tractable while retaining the same filtering and downsampling.
The design matrix therefore cannot be built at the rate at which the laser noise actually occurs, which raises the question of how the filtering, downsampling and interpolation should be represented in the design matrix.

Two further factors arise with a realistic data model:
First, at the telemetry rate the fractional delays must be interpolated close to the Nyquist frequency, where Lagrange \citep{shaddock_postprocessed_2004} and cosine-sum \citep{staab_optimal_2025} interpolators lose accuracy, and their finite kernels mean data is discarded data at the edges of every chunk.
Second, the standard TDI-\(\infty\) likelihood keeps only the null space of the design matrix, which is exact only in the limit of infinite laser noise.

In this work we extend TDI-\(\infty\) to this more realistic case of filtered and downsampled data.
We first review the construction of the design matrix and its null space (Section~\ref{sec:background}), and introduce \ac{GP} interpolation as an alternative to Lagrange and cosine-sum interpolation, which includes the interpolation error in the likelihood (Section~\ref{sec:interpolation}).
We then set up the filtered and downsampled data model, including the on-board anti-alias filter, with two models (Section~\ref{sec:realistic}), and define a likelihood that keeps both the null space and the column space of the design matrix, to test how much signal information is lost when the column space is discarded (Section~\ref{sec:likelihood}).
After describing the simulated data (Section~\ref{sec:data}), we compare the interpolators on how accurately they reproduce fractional delays and how much residual laser noise they leave (Section~\ref{sec:cancel}) and measure the cost of splitting the design matrix into chunks (Section~\ref{sec:chunksize}).
Finally we run parameter estimation on a galactic binary injected into \texttt{lisainstrument} \citep{bayle_unified_2023} simulations with all primary and secondary noises, over 200 injections, testing the interpolators and likelihoods with a wider filter band, short chunks and data gaps (Section~\ref{sec:pe}).

\section{\texorpdfstring{TDI-\(\infty\)}{TDI-∞} background}
\label{sec:background}

This section is a summary of the setup used in \citet{houba_robust_2024} which we follow in later sections.
\ac{LISA} is made of three spacecraft each containing two \acp{MOSA}, each facing one of the other two spacecraft.
Each \ac{MOSA} contains a telescope, an optical bench, a test mass and is associated with its own laser.
Each optical bench contains three interferometers, where each records the beatnote frequency between two lasers \citep{bayle_unified_2023}.
The science interferometer \textit{sci} (labelled \textit{isi} in \citet{houba_robust_2024}) beats the local laser with the remote laser from the distant \ac{MOSA} and contains the gravitational wave signal \citep{baghi_lisa_2026}.
The test mass interferometer \textit{tmi} beats the local laser reflected off the test mass with the laser of the adjacent \ac{MOSA}.
This measures the motion of the optical bench relative to the free-falling test mass, allowing the jitter of the optical bench to be removed.
The reference interferometer \textit{ref} beats the same two lasers as the \textit{tmi} but without reflecting off the test mass.
This measures the difference between the two lasers on a spacecraft, which is needed to combine the \ac{MOSA} measurements.
Each laser is phase modulated with the spacecraft clock signal, this creates laser sidebands and the beatnotes of these sidebands for the \textit{ref} and \textit{sci} are also recorded to cancel clock noise later.
As in \citet{houba_robust_2024} we use 24 of these measurements: 6 \textit{sci}, 6 \textit{tmi}, 3 \textit{ref} with 6 \textit{sci} sidebands and 3 \textit{ref} sidebands.
Only three \textit{ref} measurements are kept as the two on a single spacecraft compare the same two lasers and so carry the same primary noise \citep{bayle_unified_2023}.

There are two main groups for the noises in these measurements, primary and secondary.
There are 18 primary noises, which appear in more than one measurement. This contains the laser frequency noise of the 6 lasers, the jitter of the 6 optical benches and the noise of the 3 clocks and the 3 dominant sideband modulation noises \citep{bayle_unified_2023, hartwig_clock-jitter_2021}.
The secondary noises are different for each measurement so appear once, this includes the readout noise of each interferometer.
The primary goal of TDI-\(\infty\) is to cancel the primary noises \(\bm{p}\) from the data, as the laser noise in particular is orders of magnitude larger than the secondary noises and the signal \citep{colpi_lisa_2024}.

\subsection{Data model}
\label{sec:background:model}

In the idealised setup used by existing implementations \citep{vallisneri_tdi-infinity_2021, houba_robust_2024} the data can be written as
\begin{equation}
    \label{oldtdi:datamodel}
    \bm{y} = \bm{Mp} + \bm{n} + \bm{h}(\bm{\theta}),
\end{equation}
where for \(N_t\) time samples assuming everything is at the 4\,Hz telemetry rate, \(\bm{y}\) are the \(N_y = 24\) telemetry-rate phasemeter measurements, \(\bm{p}\) are the \(N_p = 18\) primary noises, \(\bm{M}\) is the design matrix that encodes the delays and maps the primary noises onto the measurements, \(\bm{n}\) are the secondary noises and \(\bm{h}(\bm{\theta})\) is the response of a measurement to some astrophysical signal with parameters \(\theta\).
Therefore \(\bm{y}, \bm{n}, \bm{h} \in \mathbb{R}^{N_yN_t}\), \(\bm{p} \in \mathbb{R}^{N_pN_t}\) and \(\bm{M} \in \mathbb{R}^{N_yN_t \times N_pN_t}\), where \(\bm{M}\) is not a square matrix.

In equation~(\ref{oldtdi:datamodel}) the \(\bm{M} \bm{p}\) terms are orders of magnitude larger than \(\bm{n}\) or \(\bm{h(\theta)}\) and need to be removed \citep{colpi_lisa_2024, tinto_time-delay_2021}.
The values of \(\bm{p}\) are unknown, however as they appear in more than one measurement via the delays \citep{tinto_time-delay_2021}, this can be modelled with \(\bm{M}\).
TDI-\(\infty\) uses the delays and model \(\bm{M}\) to cancel the primary noise, this needs an accurate \(\bm{M}\).

\begin{figure}
    \centering
    \includegraphics[width=1.0\linewidth]{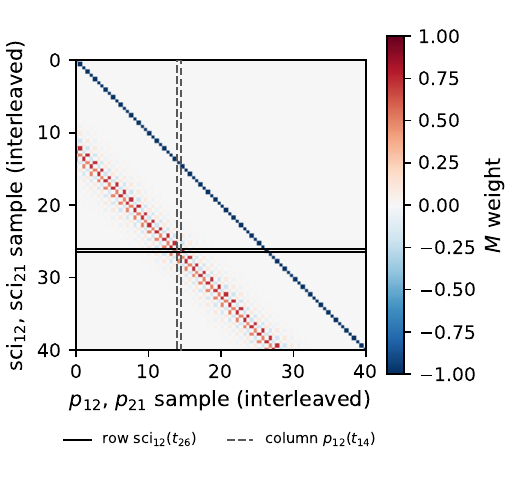}
    \caption{Structure of the design matrix \(M\), showing the mapping from the laser noises \(p_{12}, p_{21}\) (columns) to the science measurements \(\mathit{sci}_{12}, \mathit{sci}_{21}\) (rows).
        Local measurements of the laser are \(-1\) (blue) and delayed measurements are positive and interpolated.
        The solid horizontal line shows which two noises contribute to a measurement, a local measurement and an interpolated delayed measurement.
        The vertical dashed line highlights a column where there are two measurements of a particular primary noise at the same time, which is what allows cancellation.}
    \label{fig:TDI:Mmatrices}
\end{figure}

\subsection{Design matrix}
\label{oldtdi:designmatrix}

\(\bm{M}\) has 24 rows and 18 columns per time sample, relating the 18 primary noises to the 24 measurements.
The rows of \(\bm{M}\) are built using the all-in-one measurement equations of
\citet{houba_robust_2024}, however we label the inter-spacecraft interferometer measurements \textit{sci} rather than \textit{isi} in accordance with the \ac{LISA} science conventions document \citep{baghi_lisa_2026}.
The test mass and reference measurements compare lasers from the same spacecraft, therefore every noise enters at the same time and their rows have essentially no delays.
The science and science sideband measurements on \ac{MOSA} \(ij\) compare the local laser with the laser from \ac{MOSA} \(ji\).
The noise then enters through the Doppler delay operator \citep{bayle_unified_2023}, which acts on a time series \(x(t)\)
\begin{equation}
    \dot{\mathcal{D}}_{ij}x(t) = \left( 1 - \dot{d}_{ij}(t)\right) x\left(t - d_{ij}(t) \right),
\end{equation}
where \(d_{ij}\) is the light travel time along link \(ij\).
A row of \(\bm{M}\) for a science measurement then has two sets of entries: those at the time of the measurement, for the noises local to the \ac{MOSA}, and those at the delayed time, for the noises of the distant \ac{MOSA}.
The delayed time does not in general fall on a sample \(\bm{p}\), so the delayed entries are spread over neighbouring samples with an interpolator.
The standard choice, introduced for TDI in \citet{shaddock_postprocessed_2004} and used in \citet{houba_robust_2024}, is Lagrange interpolation.
The interpolators we use are described in Section~\ref{sec:interpolation}.
Fig.~\ref{fig:TDI:Mmatrices} shows this structure for a single science measurement and two of the laser noises.
The entries of \(-1\) in the \(p_{12}\) section are the local measurements, and the spread entries in \(p_{21}\) are the delayed and interpolated measurements.
The lower panel shows a single row of the M matrix.

\subsection{Null space}
\label{oldtdi:null}

The goal in TDI-\(\infty\) is to cancel the primary noise \(\bm{p}\) from only indirect measurements in \(\bm{y}\).
Therefore if we can find a matrix \(\bm{U}_0\) such that
\begin{equation}
    \label{eq:nullspace}
    \bm{U}_0^T \bm{M} = \bm{0},
\end{equation}
then pre-multiplying the data \(\bm{y}\) by \(\bm{U}_0^T\) would remove the primary noise \(\bm{p}\) \citep{vallisneri_tdi-infinity_2021, houba_robust_2024}.
In \citet{vallisneri_tdi-infinity_2021, houba_robust_2024} the variable \(\bm{T}\) is used for the null space such that \(\bm{TM} = 0\), our equivalent matrix is \(\bm{U}_0^T\).
Each of the primary noises enters more than one measurement, i.e.\ the local and delayed measurement.
Therefore there are more measurements than primary noises and the rows of \(\bm{M}\) cannot all be independent, so some combinations will sum to zero.
These combinations make the left null space of the matrix \(\bm{M}\) and they are the columns of \(\bm{U}_0\).

In \citet{houba_robust_2024} \(\bm{U}_0\) is found using a turnback algorithm which they compare against Scipy's \ac{SVD} based null space \citep{virtanen_scipy_2020}.
Here we use the \ac{SVD} of the design matrix as in \citet{vallisneri_tdi-infinity_2021}
\begin{equation}
    \bm{M} = \bm{U} \bm{\Sigma} \bm{V}^T ,
\end{equation}
where the columns of \(\bm{U}\) with zero singular value form \(\bm{U}_0\) and the remaining form the column space \(\bm{U}_r\).
Numerically we set a threshold below which the singular values are considered zero, this is discussed in Section~\ref{sec:cancel}.

Projecting the data with \(\bm{U}_0\) gives the TDI-\(\infty\) variables
\begin{equation}
    \bm{o}_0 = \bm{U}_0^T \bm{y}
    = \bm{U}_0^T \bm{n} + \bm{U}_0^T \bm{h}(\theta),
\end{equation}
which removes the primary noise and leaves the secondary noises and the signal \citep{vallisneri_tdi-infinity_2021, houba_robust_2024}.
In \citet{houba_robust_2024} the likelihood is built from \(\bm{o}_0\), we give this and its extension to the column space in Section~\ref{sec:likelihood}.

When the delays are not on integer samples in the laser noise axis of the design matrix, we need to interpolate between the times.
The standard approach, introduced for TDI in \citet{shaddock_postprocessed_2004}, uses Lagrange interpolation described in Section~\ref{sec:interpolation:lagrange}.


\section{Interpolation}
\label{sec:interpolation}

To compute the non integer delays, one needs to interpolate over previous times.
Here we compare three different interpolation methods, Lagrange \citep{shaddock_postprocessed_2004, bayle_unified_2023}, cosine-sum \citep{staab_optimal_2025} and Gaussian processes \citep{rasmussen_gaussian_2005, moore_improving_2016}.

\subsection{Lagrange}
\label{sec:interpolation:lagrange}
For some delay time \(\tau = n + \mu\), where \(n\) is an integer number of samples and \(0 \le \mu < 1\) is the fractional part, the time series \(s\) is interpolated with
\begin{equation}
    \hat{s}(\tau) = \sum_{k=-K}^K w_k(\mu)\, s(n+k),
\end{equation}
where \(K\) is the half width of the interpolation kernel. The Lagrange
weights are
\begin{equation}
    w_k(\mu) = \prod_{\substack{j=-K \\ j \neq k}}^{K} \frac{\mu - j}{k - j},
\end{equation}
each of which is the polynomial that is one at sample \(k\) and zero at the
other samples, evaluated at \(\mu\) \citep{shaddock_postprocessed_2004}.
Whilst the accuracy of this method can be tuned by selecting an appropriate order for the interpolation, any errors from the Lagrange interpolator are not accounted for in the output.
If a large number of taps are needed then there is also data loss where data is trimmed at the edges to remove the edge effects of the interpolation kernel.
These are the edges of each chunk rather than any gaps, the gaps we consider are in the telemetry samples, not in the noise itself, so the noise samples the kernel reads are complete across the gap and not affected by it.

\subsection{Cosine sum}
\label{sec:interpolation:cosine}

The cosine-sum interpolation was introduced in \citet{staab_optimal_2025}.
It has the same form as the Lagrange interpolator, with weights
\begin{equation}
    w_k(\mu) = \sum_{n=0}^{N-1} a_n \cos\!\left(\frac{2\pi n (k - \mu)}{N}\right),
\end{equation}
where \(N\) is the number of taps and \(k\) covers the surrounding \(N\) samples of the interpolated point.
The coefficients \(a_n\) are chosen in \citet{staab_optimal_2025} so that the interpolation residual stays within a stated requirement below 1\,Hz and keeps continuous first derivatives.
We use their 22 coefficients without refitting.

\subsection{Gaussian process}
\label{sec:interpolation:gp}
An alternative to Lagrange interpolation is a \ac{GP}, which propagates the interpolation error and uses the full data segment without truncation at the edges.
A \ac{GP} is defined by a kernel function \(k(\tau)\), which we choose using a model of the noise being interpolated.
Since we are interpolating stationary laser noise, we define the kernel as its auto-covariance.
For a two-sided power spectral density \(S(f)\), the kernel is
\begin{equation}
    \label{eq:interp:gpkernel}
    k(\tau) = \int_{-\infty}^{\infty} S(f)\, e^{2\pi i f \tau} \, \mathrm{d}f ,
\end{equation}
which is the inverse Fourier transform of the noise spectrum
\citep{rasmussen_gaussian_2005}.
The laser noise itself is white, however we interpolate that noise after the on-board
anti-alias filter, so for a noise variance \(\sigma^2\) and filter \(F\), \(S(f) = \sigma^2|F(f)|^2\) and the kernel is the autocorrelation of the
filter impulse response, using the parameters described in Section~\ref{sec:data:aafilter}.

We interpolate a time series \(s(t_i)\) described by a noise process \(S(f)\) to a target time series \(z(t_\alpha)\). Note that the number of samples in both the time series \(s\) and \(z\) need not be equal. Latin time-indices correspond to the original time-series, Greek time-indices the new interpolated process.
The \ac{GP} uses three covariances between the samples and targets
\begin{equation}
    \begin{split}
        k_{ss}(t_i, t_j)          & =  k(t_i - t_j),         \\
        k_{sz}(t_\alpha, t_j)     & =  k(t_\alpha - t_j),    \\
        k_{zz}(t_\alpha, t_\beta) & = k(t_\alpha - t_\beta).
    \end{split}
\end{equation}
In the following all vectors are over the sampled times so \((\bm{K})_{ij} = K_{ij} = k_{ss}(t_i, t_j)\) and \((\bm{k}(t_\alpha))_j = k_j(t_\alpha) = k_{sz}(t_\alpha, t_j)\).
This then gives us the posterior mean of a \ac{GP} \citep{rasmussen_gaussian_2005}
\begin{equation}
    \hat{z}(t_\alpha) = \bm{k}(t_\alpha)^T \bm{K}^{-1} \bm{s},
\end{equation}
so we can write the weights \(\bm{w}(t_\alpha) = \bm{k}^T(t_\alpha) \bm{K}^{-1}\) as the interpolation coefficients that populate the corresponding row of the telemetry-rate design matrix \(\bm{M}\), in place of the Lagrange weights.

Unlike the Lagrange interpolator, the \ac{GP} posterior also provides the covariance of the interpolation error \citep{moore_improving_2016}.
This \ac{GP} posterior covariance is defined by
\begin{equation}
    C_\epsilon(t_\alpha, t_\beta) = k_{zz}(t_\alpha, t_\beta) - \bm{k}^T(t_\alpha) \bm{K}^{-1} \bm{k}(t_\beta).
\end{equation}
\(C_\epsilon\) is a full covariance over target times and between measurements that share the delayed laser, while the errors of the lasers are independent in this setup.

\section{Realistic filtered data model}
\label{sec:realistic}

The idealised model in equation~(\ref{oldtdi:datamodel}) leaves out the on-board data processing.
On the spacecraft, each measurement is made with the 80\,MHz on-board rate \citep{bayle_unified_2023} as a sum of the local and delayed primary noise terms, the secondary noises and the signal.
It is then filtered with the anti-alias filter and downsampled to the 4\,Hz telemetry rate \citep{bayle_unified_2023, staab_laser_2023}.
This model is built into the data description in \citet{houba_robust_2024} but not used in the building of the \(\bm{M}\) matrix.
In the form of equation~(\ref{oldtdi:datamodel}) the model for a chunk of data can be written
\begin{equation}
    \begin{split}
        \label{eq:realistic:model}
        \bm{y} & = \bm{S}\bm{F}\left(\bm{M}_\phi \bm{p}_{\phi} + \bm{n}_{\phi} + \bm{h}_{\phi}(\theta) \right) \\
               & = \bm{P} + \bm{S}\bm{F} \bm{n}_{\phi} + \bm{S}\bm{F} \bm{h}_{\phi}(\theta),
    \end{split}
\end{equation}
where \(\bm{M}_\phi\) is the design matrix and subscript \(\phi\) refers to its sample rate at 80\,MHz, \(\bm{F}\) applies the anti-alias filter to each measurement across time, with the filter taps \(c_l\) which have \(L\) taps at the on-board sample interval.
\(\bm{S}\) downsamples by keeping samples at the telemetry times and \(\bm{P} = \bm{S}\bm{F}\bm{M}_\phi \bm{p}_{\phi}\).
The filter also introduces a group delay to the measurements, however, this is corrected on the ground and is assumed to be a fixed offset here so does not change the combinations that cancel laser noise, so we omit it here and account for it later by shifting the signal template in Sec.~\ref{sec:data:full}.

We can then write out the log likelihood in terms of the time series measurements \(\bm{y}\).
We assume here a Gaussian likelihood and that the primary \(\bm{p}_{\phi}\) and secondary noise \(\bm{n}_\phi\) are zero mean and independent and with a known covariance.
The time series likelihood is then
\begin{equation}
    \label{eq:realistic:likelihood:full}
    \begin{split}
        \log p(\bm{y} \mid \theta) = &-\frac{1}{2} \left(\bm{y} - \bm{SF} \bm{h}_{\phi}(\theta)\right)^T \bm{C}_y^{-1} \left(\bm{y} -\bm{SF} \bm{h}_{\phi}(\theta)\right) \\
        &+ \text{const.},
    \end{split}
\end{equation}
where the measurement covariance \(\bm{C}_y\) is defined by
\begin{equation}
    \label{eq:realistic:Cy:full}
    \bm{C}_y = (\bm{SF}) \left(\bm{M}_{\phi} \bm{C}_{p_{\phi}} \bm{M}_{\phi}^T + \bm{C}_{n_{\phi}} \right)(\bm{SF})^T.
\end{equation}
Here \(\bm{C}_{p_{\phi}} = \mathbb{E}[\bm{p}_\phi \bm{p}_{\phi}^T]\) is the primary noise covariance and \(\bm{C}_{n_{\phi}} = \mathbb{E}[\bm{n}_\phi \bm{n}_{\phi}^T] \) is the secondary noise covariance.

\subsection{Telemetry rate model}

In practice \(\bm{M}_\phi\) is too large to build at the 80\,MHz rate, therefore we write the model at the telemetry rate.
This means that we must move \(\bm{SF}\) through the delays in \(\bm{M}_\phi\), however these do not commute.
For the Doppler delay operator \(\dot{\mathcal{D}}\), \citet[eq.~17]{staab_laser_2023} split the error of swapping the delay with the filtering and downsampling into two parts
\begin{equation}
    \label{eq:realistic:flexfilt}
    \left[\bm{SF}, \dot{\mathcal{D}} \right] = \underbrace{\left[\bm{S}, \dot{\mathcal{D}} \right] \bm{F}}_{\text{aliasing}} + \underbrace{\bm{S} \left[\bm{F}, \dot{\mathcal{D}} \right]}_{\text{flexing-filtering}}.
\end{equation}
The first term is aliasing, where the power the filter leaves above the 2\,Hz telemetry Nyquist frequency is folded back into the band by \(\bm{S}\).
The second term is the `flexing filtering', where the delay changes over the length of the filter, so the filter and delay do not commute \citep{bayle_effect_2019}.
In \citet{harer_mitigation_2025}, the filter is corrected for by applying a modified delay operator  to the filtered and downsampled noise \(\bm{SFp}\) (their Eq. 22)
\begin{equation}
    \label{eq:realistic:filtercorrect}
    \hat{\mathcal{D}} = \bm{F} \dot{\mathcal{D}} \bm{F}^{-1}.
\end{equation}
We take this and apply it to TDI-\(\infty\), where as the design matrix is built from delayed and local terms, this same correction would apply to \(\bm{M}\).
At the telemetry rate the primary noise term becomes
\begin{equation}
    \label{eq:realistic:primarynoise}
    \bm{P} \approx \hat{\bm{M}}_4 \bm{SFp}_\phi, \quad \hat{\bm{M}}_4 = \bm{F}_4\bm{M}_4 \bm{F}_4^{-1},
\end{equation}
where \(\bm{M}_4\) is the design matrix of Section~\ref{oldtdi:designmatrix} at the telemetry rate, \(\bm{F}_4\) is the filter resampled to 4\,Hz. Both \(\bm{F}\) and \(\bm{F}_4\) apply the same filter to each time series separately, so their size is set by what they act on.
We cannot form \(\bm{F}_4^{-1}\) directly, however, we will never need to.
The two models below are two ways of applying \(\hat{\bm{M}}_4\).
Model 1 in Section~\ref{sec:realistic:model1} splits \(\hat{\bm{M}}_4\) into an interpolation followed by the filter and handles flexing filtering but not aliasing and works with all interpolators from Section~\ref{sec:interpolation}.
Model 2 in Section~\ref{sec:realistic:model2} builds all of \(\hat{\bm{M}}_4\) into the \ac{GP} weights and can include both error terms, however is only applied to the \ac{GP} interpolation.
In both models each row of the design matrix combines local terms, which are the samples \(\bm{s}\) at the telemetry times, and delayed terms \(\bm{z}\), which are interpolated from \(\bm{s}\).
The two models differ only in how they define \(\bm{s}\) and \(\bm{z}\).

\subsection{Model 1: delays before filter}
\label{sec:realistic:model1}

In Model 1 we group the operators in equation~(\ref{eq:realistic:primarynoise}) as
\begin{equation}
    \label{eq:realistic:lagrange:approx}
    \bm{P} \approx \left(\bm{F}_4 \bm{M}_4 \right) \left(\bm{F}_4^{-1} \bm{S} \bm{F} \bm{p}_\phi \right) = \bm{F}_4 \bm{M}_4 \bm{s}_1
\end{equation}
where \(\bm{s}_1 = \bm{F}_4^{-1} \bm{S} \bm{F} \bm{p}_\phi\) is the noise at the telemetry rate with the filter undone.
This effectively assumes that \(\bm{s}_1\) is band limited to 2\,Hz, and there is no aliasing back into the band, so this model does not account for any aliasing.
The unknowns in this case are \(\bm{s}_1\) so \(\bm{F}_4^{-1}\) is never formed and the delayed terms are \(\bm{z}_1 = \dot{\mathcal{D}}\bm{s}_1\).
The interpolation weights in \(\bm{M}_4\) form the noise at the delayed times as combinations of \(\bm{s}_1\), \(\bm{F}_4\) is then applied to the result.
As the delays are applied before the filter, this handles the flexing filtering of equation~(\ref{eq:realistic:flexfilt}).

The Lagrange and cosine-sum models fill \(\bm{M}_4\) with their interpolation weights from Sections~\ref{sec:interpolation:lagrange} and~\ref{sec:interpolation:cosine}.
The \ac{GP} can be used here also as \(\bm{s}_1\) is white and band limited to 2\,Hz, therefore the kernel of equation~
(\ref{eq:interp:gpkernel}) is a sinc,
\begin{equation}
    k(t) = \sigma_p^2 \;\mathrm{sinc}(t/\Delta t),
\end{equation}
where \(\Delta t\) is the telemetry sample spacing and \(\sigma_p^2\) is the variance of a sample.
The samples are then uncorrelated and \(\bm{K}=\sigma_p^2 \bm{I}\) so the weights are just a sinc function.
As \(\bm{s}_1\) and \(\bm{z}_1\) are the same unfiltered noise, all three covariances from Section~\ref{sec:interpolation:gp} are the same kernel read at different lags.

\subsection{Model 2: filter in \texorpdfstring{\ac{GP}}{GP} kernel}
\label{sec:realistic:model2}

In Model 2 we model the whole \(\hat{\bm{M}}_4\) and apply it directly to the filtered and downsampled noise,
\begin{equation}
    \bm{P} \approx \hat{\bm{M}}_4 (\bm{SF} \bm{p}_\phi) = \hat{\bm{M}}_4 \bm{s}_2.
\end{equation}
In this case the unknowns are the samples \(\bm{s}_2 = \bm{SFp}_\phi\) and the \ac{GP} weights in \(\hat{\bm{M}}_4\) write the filtered delayed noise, \(\bm{z}_2 \approx \hat{\mathcal{D}}_4 \bm{s}_2\), with \(\hat{\mathcal{D}}_4 = \bm{F}_4 \dot{\mathcal{D}} \bm{F}_4^{-1}\) as combinations of them.
Reading the kernel of Section~\ref{sec:interpolation:gp} at the delayed time, \(k_{sz}(t_\alpha, t_j) = k(t - d - t_j)\), handles neither term of equation~(\ref{eq:realistic:flexfilt}).
Here we include the effects of aliasing through the kernel integrals and the flexing filtering through the filtered delayed terms, then build the \ac{GP} from the resulting kernels.

\subsubsection{Aliasing}
To account for aliasing, the kernel in equation~(\ref{eq:interp:gpkernel}) is built over the full on-board band.
Reading this kernel at samples separated by \(\Delta t\) then includes the power above 2\,Hz that \(\bm{S}\) folds in.
The part that the \ac{GP} cannot predict from the aliased samples is included in \(\bm{C}_\epsilon\).
We build the kernels from the on-board 16\,Hz simulation rate, in practice one would have to select a rate for this lower than the 80\,MHz such as this, however, we do not investigate that here.

\subsubsection{Flexing filtering}

Reading the kernel at a delayed time treats \(\hat{\mathcal{D}}\) as \(\dot{\mathcal{D}}\), which only holds for constant delay.
For changing delays, we expand in the rate of change of the delay as in \citet{staab_laser_2023, harer_mitigation_2025}.
To do this we write the samples \(\bm{s}_2 = \bm{SFp}_\phi\) and a delayed term as sums over the filter taps.
The samples are filtered noise at the telemetry times,
\begin{equation}
    s_2(t_j) = \sum_{l=0}^{L-1} c_l p(t_j - t_l),
\end{equation}
where \(t_l\) are the filter lags at the on-board sample interval, \(L\) is the number of filter taps.
A delayed term is an element of \(\bm{z}_2 = \bm{S}\bm{F}\dot{\mathcal{D}}\bm{p}_\phi \approx \hat{\mathcal{D}}\,\bm{s}_2\), which is the filtered delayed noise that \(\hat{\bm{M}}_4\) predicts.
On the spacecraft each delayed term is
\begin{equation}
    z_2(t) = \sum_{l=0}^{L-1} c_l \left(1 - \dot d(t - t_l)\right)p(t - t_l - d(t - t_l))
\end{equation}
If the delays were constant \(z_2(t) = s_2(t - d)\) and the delay could be moved outside the filter.
We assume however that the delay changes linearly over the filter length, and expand it to first order about the centre of the filter, which is a group delay \(t_g\) before \(t\)
\begin{equation}
    d(t - t_l) \approx d_c + \dot{d}_c(t_g - t_l),
\end{equation}
where \(d_c\) and \(\dot d_c\) are the delay and its time derivative at the filters center \(t - t_g\).
We assume the \(\ddot d\) terms are small, so the Doppler factor term becomes \((1 - \dot d_c)\), giving
\begin{equation}
    z_2(t) \approx (1 - \dot d_c)\sum_{l=0}^{L-1} c_l p(t - d_c - t^{\prime}_l), \quad t^\prime_l = t_l + \dot d_c(t_g - t_l).
\end{equation}
This means that the delay \(d_c\) does not depend on the filter taps and acts after the filter.
The change of the delay across the filter has moved into the lags.
Therefore \(z_2\) is the noise passed through a second filter with the same taps but lags \(t_l^\prime\) instead of \(t_l\) and read at \(t_\alpha = t - d_c\).
The two filters are then defined by
\begin{equation}
    \begin{split}
        F(f)   & = \sum_l c_l e^{-2\pi i f t_l},                       \\
        F_z(f) & = (1 - \dot d_c)\sum_l c_l e^{-2\pi i f t_l^\prime} ,
    \end{split}
\end{equation}
where the term \(1 - \dot d_c\) is in \(F_z\) so that \(z_2\) is the noise passed through \(F_z\) and read at \(t_\alpha\).
To first order in \(\dot{d}_c\) \(F_z\) reproduces the flexing filtering term of \citet[eq.~18]{staab_laser_2023}, which \citet[eq.~24]{harer_mitigation_2025} use to correct for the delay.
Where \citet{harer_mitigation_2025} design an interpolation kernel for this correction, here it is contained in the \ac{GP} through \(F_z\).

\subsubsection{GP kernels}

The samples pass through the filter \(F\) and are read at times \(t_j\) and a delayed term passes through \(F_z\) and is read at \(t_\alpha\).
The covariance between \(s\) and \(z\) then contains the power spectrum \(S_p\) and filters with the two different filters.
The three kernels for the \ac{GP} are therefore defined by
\begin{equation}
    \begin{split}
        k_{ss}(t_i, t_j)          & = \int S_p |F|^2 e^{2\pi i f (t_i - t_j)} \mathrm{d}f,                                 \\
        k_{sz}(t_\alpha, t_j)     & = \int S_p F_zF^* e^{2\pi i f (t_\alpha - t_j)} \mathrm{d}f,                           \\
        k_{zz}(t_\alpha, t_\beta) & = \int S_p F_{z, \alpha}F_{z, \beta}^*  e^{2\pi i f (t_\alpha - t_\beta)} \mathrm{d}f,
    \end{split}
\end{equation}
where the integrals are over the full frequency band, and \(F_{z, a}\) is taken at time \(t_a\).
When \(\dot{d}_c = 0\), \(F_z = F\) and this reduces to equation~(\ref{eq:interp:gpkernel}).
The weights and covariance then follow that from Section~\ref{sec:interpolation:gp}.
The filter suppresses the stopband power by a factor of about \(10^{18}\), which is below double precision.
This makes the condition number of \(\bm{K}\) \(\sim 10^{18}\) and when inverting it amplifies the float64 errors.
We therefore form the kernel and \(\bm{C}_\epsilon\) and solve for the \ac{GP} weights in 128-bit precision with \texttt{python-flint} \citep{flintlibpython-flint}.

\subsection{Likelihood at telemetry rate}
\label{sec:realistic:likelihood}

The likelihood from equation~(\ref{eq:realistic:likelihood:full}) can then be rewritten at the telemetry rate
\begin{equation}
    \label{eq:realistic:likelihood}
    \log p(\bm{y} \mid \theta) \propto -\frac{1}{2} \left(\bm{y} - \bm{F}_4 \bm{h}_{4}(\theta)\right)^T \bm{C}_y^{-1} \left(\bm{y} -\bm{F}_4 \bm{h}_{4}(\theta)\right).
\end{equation}
The measurement covariance \(\bm{C}_y\) is defined by
\begin{equation}
    \label{eq:realistic:Cy}
    \bm{C}_y = \bm{M} \bm{C}_{s} \bm{M}^T +  \tilde{\bm{C}}_{n} ,
\end{equation}
where we combine the secondary noise covariance with the \ac{GP} error covariance
\begin{equation}
    \label{eq:likelihood:ctilde}
    \tilde{\bm{C}}_n = \bm{F}_4 \bm{C}_{n_4}\bm{F}_4^T + \bar{\bm{C}}_\epsilon,
\end{equation}
where \(\bar{\bm{C}}_\epsilon\) is the covariance from the \ac{GP} and is only included for the matrices that use \ac{GP} interpolation.
In both models we can write the primary noise as \(\bm{M}\bm{s}\), where \(\bm{s}\) are the unknown samples with covariance \(\bm{C}_s\).
In Model 1, \(\bm{M} = \bm{F}_4 \bm{M}_4\) and \(\bm{s} = \bm{s}_1\) with each primary noise given its spectrum from Table~\ref{tab:noises}, which is white for laser noise.
In Model 2, \(\bm{M} = \hat{\bm{M}}_4\) and \(\bm{s} =\bm{s}_2\) with the same spectra multiplied by the filter \(| F | ^2\).
In Model 1, \(\bar{\bm{C}}_\epsilon = 0\) for the Lagrange and cosine-sum and \(\bar{\bm{C}}_\epsilon = \bm{F}_4 \bm{C}_\epsilon \bm{F}_4^{T}\) for the \ac{GP}.
In Model 2, \(\bar{\bm{C}}_\epsilon = {\bm{C}}_\epsilon\) is the \ac{GP} covariance itself.

\section{Likelihood definition}
\label{sec:likelihood}

The TDI-\(\infty\) likelihood is usually written in terms of the null space of the design matrix.
The laser noise is marginalised with an infinite variance prior, the remaining column space directions are dropped entirely \citep{vallisneri_tdi-infinity_2021}.
This is a good approximation when the laser noise is large, however, the discarded directions can still carry information about the signal.
In later sections we show how this matters for data split into short chunks or with gaps, and in Appendix~\ref{app:threshold} how it depends on the threshold that separates the two spaces.
In this section we keep both spaces and write out the likelihood for each. The singular values of the design matrix set how much weight the column space is given.

We start with the likelihood definition in equation~(\ref{eq:realistic:likelihood}), whiten all variables to the secondary noises and split the likelihood into null and column space components.

\subsection{Whitening}

To simplify the likelihood computation, we whiten the components of the likelihood by the secondary noise plus any interpolation error \(\tilde{\bm{C}}_n\), so that after laser noise cancellation the remaining noise is white.
The data, signal template and design matrix are whitened by
\begin{equation}
    \begin{split}
        \bar{\bm{y}} & = \tilde{\bm{C}}_n^{-1/2} \bm{y},              \\
        \bar{\bm{h}} & = \tilde{\bm{C}}_n^{-1/2} \bm{F}_4 \bm{h}_{4}, \\
        \bar{\bm{M}} & = \tilde{\bm{C}}_n^{-1/2} \bm{M}.
    \end{split}
\end{equation}
We compute  \(\tilde{\bm C}_n^{-1/2}\) from the eigen decomposition of a known \(\tilde{\bm C}_n\), keeping only the \(N_w\) eigenvectors whose eigenvalues are above a fraction of the largest. 
This is then not a true inverse square root, but an \(N_w \times N_y\ell\) projection onto the retained eigenvectors, so \(\bar{\bm y}\), \(\bar{\bm h}\) and \(\bar{\bm M}\) each have \(N_w\) rows.  
The anti-alias filter suppresses the secondary noise in its stop band by more than double precision, so the smallest eigenvalues are round off and whitening to them would amplify numerical noise.
In Appendix~\ref{app:cncut} we show that we can cut to a threshold of \(10^{-10}\) of the largest eigenvalue.

\subsection{Singular value covariance}

We take the singular value decomposition of the whitened design matrix, \(\bar{\bm{M}} = \bm{U} \bm{\Sigma} \bm{V}^T\), where \(\bar{\bm{M}}\) has \(N_w\) rows, from the whitening, and rank \(r\).
We split the left singular vectors into the column space \(\bm{U}_r\) and null space \(\bm{U}_0\), \(\bm{U} = [\bm{U}_r, \bm{U}_0]\).
Projecting the whitened data onto these subspaces gives
\begin{equation}
    \begin{split}
        \bm{o}_0 & = \bm{U}_0^T \bar{\bm{y}}, \\
        \bm{o}_r & = \bm{U}_r^T \bar{\bm{y}},
    \end{split}
\end{equation}
the null and column space projections.
Together these are the full projection of the whitened data,
\begin{equation}
    \bm{o} = \bm{U}^T \bar{\bm{y}} = \begin{bmatrix} \bm{o}_r \\ \bm{o}_0 \end{bmatrix},
\end{equation}
which are the TDI-\(\infty\) observables, with the standard laser-free variables in \citet{vallisneri_tdi-infinity_2021, houba_robust_2024} being the
\(\bm{o}_0\) part alone.
Whitening the covariance of equation~(\ref{eq:realistic:Cy}) gives
\begin{equation}
    \label{eq:Cybar}
    \bm{C}_{\bar{y}} = \bar{\bm{M}} \bm{C}_{s} \bar{\bm{M}}^T + \bm{I},
\end{equation}
so the covariance of the projected data \(\bm{o}\), which we write \(\bm{C}_o\), is
\begin{equation}
    \begin{split}
        \bm{C}_o & = \bm{U}^T \bm{C}_{\bar{y}} \bm{U}                                                                                                                           \\
                 & = \bm{U}^T \left(\bar{\bm{M}} \bm{C}_{s} \bar{\bm{M}}^T + \bm{I} \right) \bm{U}                                                                              \\
                 & = \underbrace{\bm{U}^T \bar{\bm{M}} \bm{C}_{s} \bar{\bm{M}}^T \bm{U}}_{\text{primary noise}} + \underbrace{\bm{U}^T \bm{I}\bm{U}}_{\text{secondary noise}} .
    \end{split}
\end{equation}
Using \(\bar{\bm{M}} = \bm{U} \bm{\Sigma V}^T\) we can write \(\bm{U}^T \bar{\bm{M}} = \bm{\Sigma} \bm{V}^T\) and since \(\bm{U}\) is unitary \(\bm{U}^T \bm{I} \bm{U}= \bm{I}\), giving
\begin{equation}
    \bm{C}_o = (\bm{\Sigma V}^T) \bm{C}_{s} (\bm{V \Sigma}^T) + \bm{I} .
\end{equation}
\(\bm{\Sigma}\) has non-zero entries only on the leading \(r\) diagonal elements. Writing
\(\bm{\Sigma}_r\) for the \(r \times r\) block that acts on the column space, the remaining
\(N_w - r\) rows of \(\bm{\Sigma}\) are zero, and \(\bm{C}_o\) is block diagonal,
\begin{equation}
    \label{eq:covo}
    \bm{C}_o = \begin{bmatrix}
        \bm{\Sigma}_r (\bm{V}_r^T \bm{C}_{s} \bm{V}_r) \bm{\Sigma}_r^T + \bm{I}_r & \bm{0}         \\
        \bm{0}                                                                    & \bm{I}_{N_w-r}
    \end{bmatrix} .
\end{equation}
The null space block is the identity matrix as no laser noise should reach \(\bm{o}_0\) and only the whitened secondary noise is left.
The column space has the primary noise, including laser noise, weighted by the singular values.

\subsection{Split likelihood}

In equation~(\ref{eq:covo}) the covariance of the projected data is block diagonal, the likelihood of equation~(\ref{eq:realistic:likelihood}) then separates into null space and column space parts,
\begin{equation}
    \log p(\bar{\bm{y}} \mid \theta) = \log p(\bm{o}_0 \mid \theta) + \log p(\bm{o}_r \mid \theta) .
\end{equation}
The null space part has covariance \(\bm{I}\) of size \(N_w - r\),
\begin{equation}
    \label{likelihood:null}
    \log p(\bm{o}_0 \mid \theta) = -\frac{1}{2} \left\lVert \bm{o}_0 - \bm{U}_0^T \bar{\bm{h}} \right\rVert^2 ,
\end{equation}
and the column space part has covariance \(\bm{C}_r = \bm{\Sigma}_r (\bm{V}_r^T \bm{C}_{s} \bm{V}_r) \bm{\Sigma}_r^T + \bm{I}_r\),
\begin{equation}
    \label{likelihood:column}
    \begin{split}
        \log p(\bm{o}_r \mid \theta) & = -\frac{1}{2} \left(\bm{o}_r - \bm{U}_r^T \bar{\bm{h}} \right)^T \bm{C}_r^{-1}\left(\bm{o}_r - \bm{U}_r^T \bar{\bm{h}} \right) \\
                                     & -\frac{1}{2}\log\det \bm{C}_r .
    \end{split}
\end{equation}
Standard TDI-\(\infty\) \citep{vallisneri_tdi-infinity_2021} uses only the null space part \(\log p(\bm{o}_0 \mid \theta)\).
In the limit of infinite primary noise, \(\bm{C}_{s} \to \infty\), the column space covariance \(\bm{C}_r \to \infty\) and \(\log p(\bm{o}_r \mid \theta)\) no longer depends on \(\theta\), recovering the null-space likelihood.
When the laser noise is finite the column space part contributes, with the singular values \(\bm{\Sigma}_r\) setting the scale of its covariance.

The null space likelihood of equation~(\ref{likelihood:null}) does not depend on the assumed
primary noise level for the Lagrange and cosine-sum, because \(\bm{U}_0\) is set by \(\bar{\bm{M}}\) alone.
The \ac{GP} scales \(\bm{C}_\epsilon\) by the primary noise variance.
The column space term \(\bm{C}_r\) depends directly on \(\bm{C}_{s}\), so the wrong primary noise level gives the column space directions the wrong weight.


\section{Simulated data}
\label{sec:data}

In the tests in the following sections we use two main data setups. The first is a simple single channel setup for testing the interpolators' accuracy.
The second is a simulation of the full constellation used for the full design matrix tests (Section~\ref{sec:cancel}), chunk size tests (Section~\ref{sec:chunksize}) and parameter estimation tests (Section~\ref{sec:pe}).
The configuration files and code used to run all the tests can be found in the project repository \citep{bayley_tdiinf_zenodo_2026}.

\subsection{Anti-aliasing filter}
\label{sec:data:aafilter}

All setups use the same anti-alias filter design from \citet{staab_laser_2023}, which is a Kaiser FIR filter with a passband edge at 0.1\,Hz, stopband edge at 1\,Hz or 2\,Hz and 181\,dB attenuation. 
This is made with \texttt{lisainstrument}'s \texttt{make\_fir\_causal\_kaiser} \citep{bayle_unified_2023} using \textit{scipy}'s \texttt{kaiserord} \citep{virtanen_scipy_2020}.
We use two different stopband edges, one at the Nyquist of 2\,Hz and one at 1\,Hz.
The filter is applied to the data at the 16\,Hz simulation rate and the models use the same filter resampled to the 4\,Hz telemetry rate as \(F_4\).
We simulate the data at 16\,Hz rather than the full 80\,MHz on-board rate, 16\,Hz is the \texttt{lisainstrument} default \citep{staab_laser_2023}.

\subsection{Single channel test}
\label{sec:data:single}

To test the three interpolators of Section~\ref{sec:interpolation}, we simulate one channel of white noise with unit variance for 4000\,s with sample rate of 16\,Hz.
It is then filtered with the filter from Section~\ref{sec:data:aafilter} with the stopband at 2\,Hz and downsampled to 4\,Hz.
The interpolators then resample the noise back to 16\,Hz to compare with the filtered 16\,Hz data in Section~\ref{sec:cancel}.

\subsection{Full constellation}
\label{sec:data:full}

The full constellation is simulated with \texttt{lisainstrument} v2.1.0 \citep{bayle_unified_2023} at a 16\,Hz sample rate, with the noise levels defined in Table~\ref{tab:noises}.
All four primary noises are on, giving the full 18 columns per telemetry sample.
The secondary noises are the \ac{OMS} noise and the test-mass noise, all other secondary noises (ranging, angular jitter, tilt-to-length) are off.
The spacecraft follow the ESA science orbits \citep{martens_trajectory_2021, bayle_lisaorbits_2025}, so arm lengths vary with time.
The lasers are locked in the ``six'' configuration, so the lasers are independent.
The lasers have a central frequency of \(\nu_0 = 2.816\times10^{14}\)\,Hz, and its noise is white. 
The data is filtered with the same filters described in Section~\ref{sec:data:aafilter} and downsampled to 4\,Hz.
The group delay of the filter is not removed from the data as in Sec.~\ref{sec:realistic}, instead the signal template is shifted by \(t_g\) to align with the data.
The whole data segment is 500 samples or 125\,s long (1000 samples for the chunk size test of Section~\ref{sec:chunksize}), and is then further split into various chunking sizes.

For parameter estimation a galactic binary signal is added before the filter stage so it passes the same filtering and downsampling as the noise, signals are defined in Section~\ref{sec:data:signal}.

\begin{table}
    \centering
    \caption{Simulation and injection settings. The top section shows the noise levels used in the simulations, which are \texttt{lisainstrument} defaults \protect\citep{bayle_unified_2023}. Test mass acceleration noise has the default spectrum with 0.4\,mHz knee and 8\,mHz break. Bottom shows the galactic binary injection priors, where \(A, f, \phi\) are drawn from uniform priors over the shown ranges and the remaining parameters are fixed to the stated values.}
    \label{tab:noises}
    \begin{tabular}{llcl}
        \hline
        Noise                       & No.                                                              & ASD                                                             \\
        \hline
        \multicolumn{3}{l}{\textit{Primary, marginalised}}                                                                                                               \\
        Laser frequency             & 6                                                                & \(30\,\mathrm{Hz}/\sqrt{\mathrm{Hz}}\)                          \\
        Optical bench jitter        & 6                                                                & \(2\times10^{-9}\,\mathrm{m}/\sqrt{\mathrm{Hz}}\)               \\
        Clock                       & 3                                                                & \(6.32\times10^{-14}/\sqrt{\mathrm{Hz}}\) at 1\,Hz              \\
        Modulation                  & 3                                                                & \(5.2\times10^{-13}/\sqrt{\mathrm{Hz}}\) at 1\,Hz               \\
        \hline
        \multicolumn{3}{l}{\textit{Secondary, not cancelled}\ \ }                                                                                                        \\
        \ac{OMS} science carrier    & 6                                                                & \(6.35\times10^{-12}\,\mathrm{m}/\sqrt{\mathrm{Hz}}\)           \\
        \ac{OMS} science sideband   & 6                                                                & \(1.25\times10^{-11}\,\mathrm{m}/\sqrt{\mathrm{Hz}}\)           \\
        \ac{OMS} test-mass carrier  & 6                                                                & \(1.42\times10^{-12}\,\mathrm{m}/\sqrt{\mathrm{Hz}}\)           \\
        \ac{OMS} reference carrier  & 6                                                                & \(3.32\times10^{-12}\,\mathrm{m}/\sqrt{\mathrm{Hz}}\)           \\
        \ac{OMS} reference sideband & 6                                                                & \(7.90\times10^{-12}\,\mathrm{m}/\sqrt{\mathrm{Hz}}\)           \\
        Test-mass acceleration      & 6                                                                & \(2.4 \times 10^{-15}\,\mathrm{m\,s}^{-2}/\sqrt{\mathrm{Hz}} \) \\
        \hline
        \hline
        Parameter                   & \multicolumn{2}{l}{Value}                                                                                                          \\
        \hline
        \multicolumn{3}{l}{\textit{Galactic binary, sampled, uniform prior range}}                                                                                       \\
        \(A\)                       & \multicolumn{2}{l}{\((2 \times 10^{-20},\, 7 \times 10^{-20})\)}                                                                   \\
        \(f\)                       & \multicolumn{2}{l}{\((0.02,\, 0.05)\,\mathrm{Hz}\)}                                                                                \\
        \(\phi_0\)                  & \multicolumn{2}{l}{\((-\pi,\, \pi)\)}                                                                                              \\
        \multicolumn{3}{l}{\textit{Galactic binary, fixed}}                                                                                                              \\
        \(\alpha\) (RA)             & \multicolumn{2}{l}{4.6511\,rad}                                                                                                    \\
        \(\delta\) (Dec)            & \multicolumn{2}{l}{\(-0.5060\)\,rad}                                                                                               \\
        \(\iota\)                   & \multicolumn{2}{l}{\(\pi/3\)\,rad}                                                                                                 \\
        \(\psi\)                    & \multicolumn{2}{l}{0.5\,rad}                                                                                                       \\
        \(\dot{f}\)                 & \multicolumn{2}{l}{0}                                                                                                              \\
        \hline
    \end{tabular}
\end{table}

\subsection{Signal injections}
\label{sec:data:signal}

The injected signals are monochromatic galactic binaries and are injected using \texttt{lisagwresponse} \citep{bayle_2025_17238548}. The amplitude \(A\), the frequency \(f\) and the initial phase \(\phi_0\) are sampled uniformly from the ranges in Table~\ref{tab:noises}, which are unrealistically loud, but chosen to make signals detectable with an \ac{SNR} over 100 for the short segments.
The remaining galactic binary parameters remain fixed for these tests.
We do not perform a search, all the chains start close to the injection parameters as we are interested in the width and calibration of the posterior rather than the sampler.

\subsection{Configurations}
\label{sec:data:configurations}

We compare five configurations in the following section, each using the 1\,Hz and 2\,Hz stopband edges of Section~\ref{sec:data:aafilter}, these are shown in Table~\ref{tab:configs}.
The full model builds \(\bm{M}_{\phi}\) of equation~(\ref{eq:realistic:model}) at the simulation rate of 16\,Hz.

\begin{table}
    \centering
    \caption{These are the TDI-\(\infty\) configurations we compare, and which of the two filter errors of equation~(\ref{eq:realistic:flexfilt}) each handles.}
    \label{tab:configs}
    \begin{tabular}{llcc}
        \hline
        Configuration     & Model                                  & Flexing filtering & Aliasing \\
        \hline
        Lagrange          & 1 (Section~\ref{sec:realistic:model1}) & yes               & no       \\
        Cosine-sum        & 1                                      & yes               & no       \\
        \ac{GP}, Model 1  & 1                                      & yes               & no       \\
        \ac{GP}, Model 2  & 2 (Section~\ref{sec:realistic:model2}) & yes               & yes      \\
        Full 16\,Hz model & --                                     & exact             & exact    \\
        \hline
    \end{tabular}
\end{table}

\section{Interpolation and laser noise cancellation}
\label{sec:cancel}

In TDI-\(\infty\) the interpolator sets the fractional delays in the design matrix \(\bm{M}\), and as the laser noise is orders of magnitude larger than the secondary noise, small interpolation errors could have a large impact on the residual laser noise in the null space.  Here we test how well the three interpolators of Section~\ref{sec:interpolation} reproduce fractional delays, and what residual laser noise they leave.

\subsection{Interpolation accuracy}
\label{sec:cancel:interp}

In the first test we use the simplified single channel data from Section~\ref{sec:data:single}, so that we can compare each interpolator's reconstruction of the filtered data.

\begin{figure}
    \centering
    \includegraphics[width=0.9\linewidth]{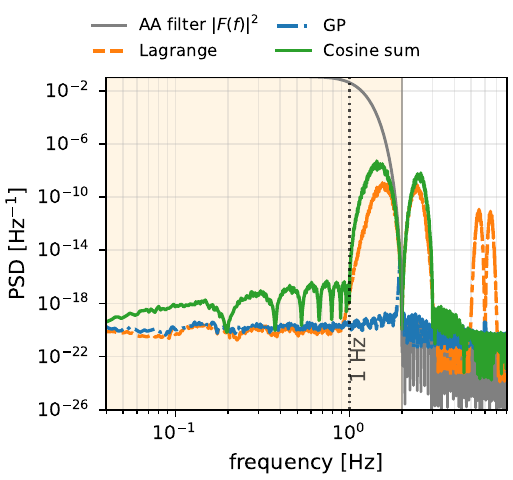}
    \caption{Frequency spectra of the difference between the 16\,Hz filtered data and the same data downsampled to the 4\,Hz telemetry rate and interpolated back to 16\,Hz for
        the three interpolation methods. The dotted line at 1\,Hz marks the passband edge over which the cosine-sum coefficients were fit. The grey curve is the anti-alias filter response \(|F(f)|^2\).}
    \label{fig:interp_accuracy}
\end{figure}

Fig.~\ref{fig:interp_accuracy} shows the spectrum of the residuals between the interpolator reconstructions and the original filtered 16\,Hz data.
The Lagrange and cosine-sum reconstructions show large peaks below and above the 2\,Hz Nyquist frequency.
These lie outside the 1\,Hz \ac{LISA} requirements band, but they are not by default filtered out before cancellation and so still contaminate the TDI-\(\infty\) null space.
The \ac{GP} is the most accurate interpolator between 1 and 2\,Hz because its kernel carries information about the anti-alias filter that produced the data.
The cosine-sum error is the largest of the three.
Its coefficients are fit over the interpolator's passband reaching only half the Nyquist, as required for \ac{LISA} \citep{staab_optimal_2025}, which on the 4\,Hz grid places the passband edge at 1\,Hz, marked on the figure.
All interpolators have a similar error  between \(10^{-4}\) and \(10^{-1}\)\,Hz.

\subsection{Full design matrix}
\label{sec:cancel:fulldesign}

\begin{figure}
    \centering
    \includegraphics[width=0.9\linewidth]{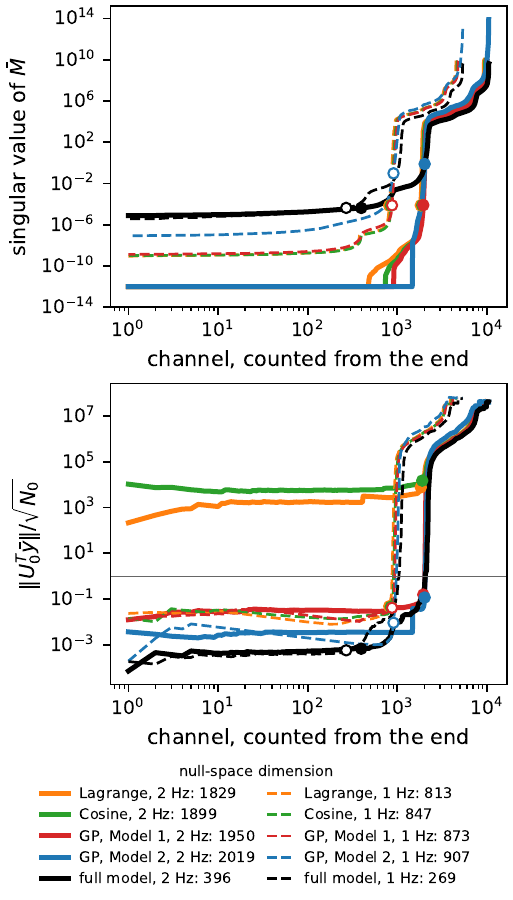}
    \caption{The singular values of the \(\bar{\bm{M}}\) matrix for the five configurations of Table~\ref{tab:configs} and two filter stop bands of 1\,Hz and 2\,Hz.
        The top panel shows the singular values plotted in reverse order.
        The lower panel shows the laser noise cancellation for each of these runs where the horizontal line at 1.0 refers to the secondary noise variance after whitening.
        The circles in both panels are the threshold used to define the null space, \(\sigma_{\rm cut}/\sigma_{\rm max} = 10^{-14}\). }
    \label{fig:singular_tdiinf}
\end{figure}

We test the laser noise cancellation using a noise only segment of full constellation data described in Section~\ref{sec:data:full}.
However in this test the secondary noises are switched off so that the projected data contains only the uncancelled primary noise.
We still whiten by \(\tilde{\bm{C}}_n\) to match that used in the later tests.
We build the \(\bm{M}\) matrix using the five configurations of Table~\ref{tab:configs} at the two filter stopband edges at 1\,Hz and 2\,Hz.
Then we compute the \ac{SVD} of the design matrix and find the  \ac{RMS} of the null space projection  \(\lVert \bm{U}_0^T \bar{\bm{y}} \rVert\).
We define the singular value threshold at \(10^{-14}\) of the largest singular value, which is the round off floor of the \ac{SVD}.
Appendix~\ref{app:threshold} shows how the null space results depend on this choice
The circles in both panels of Fig.~\ref{fig:singular_tdiinf} also show where this threshold sits.
In the top panel we can see the singular values from the \ac{SVD} of the ten runs.
The 1\,Hz runs have a smaller null space than the 2\,Hz because the filter suppresses more of the 1--2\,Hz band, so the whitening cut of Section~\ref{sec:likelihood} removes these components.
The full model has a smaller null space due to the 16\,Hz columns giving more primary noise unknowns per measurement (Section~\ref{sec:background}).
We can see that in the second panel all residuals fall below the whitened secondary noise of 1.0, except for the Lagrange and cosine-sum with a 2\,Hz cutoff.
This is due to the large interpolation errors seen in Section~\ref{sec:cancel:interp}.
The \ac{GP} however performs similarly for both cutoffs cancelling laser noise at higher frequencies near Nyquist well.

\section{Chunk sizes}
\label{sec:chunksize}

\begin{figure}
    \centering
    \includegraphics[width=1.0\linewidth]{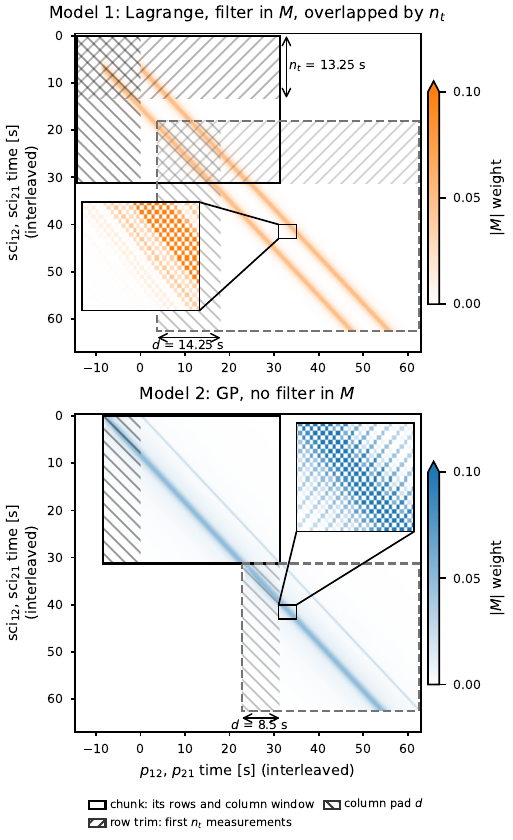}
    \caption{Two chunks of 125 samples of the design matrix showing only \(\mathit{sci}_{12}(t), \mathit{sci}_{21}(t)\) against \(p_{12}(t), p_{21}(t)\) where each has its samples interleaved in time. The top panel shows Model 1 (Section~\ref{sec:realistic:model1}) using Lagrange interpolation and an overlap of \(N_{\text{trim}}\) and the second panel shows the same for Model 2 (Section~\ref{sec:realistic:model2}) using the \ac{GP}.
        The hatched regions of the boxes show both the padding of the noise columns by \(d\) in both models and also the row trims in Model 1.
        The insets show a full resolution section of the matrix, where the interleaved structure is visible.}
    \label{fig:chunk_trim}
\end{figure}

The null space of \(\bar{\bm{M}}\) is found with a singular value decomposition, whose cost
grows as the cube of the number of measurements, so for a realistic observation \(\bm M\) cannot be factorised in one piece.
Following \citet{houba_robust_2024} we split \(\bm M\) into chunks of \(\ell\) samples along the time axis and take the null space of each.
\citet{houba_robust_2024}  use \(\ell = 500\) at 4\,Hz with a 10 per cent overlap to keep some of the noise cross correlation lost when the covariance is split, whereas we use overlap only to recover the rows removed after the filter trim in Fig.~\ref{fig:chunk_trim}.
In practice any chunk that trims rows would be overlapped, so in this section we compare Model 1 configurations with overlaps, and include non overlapped chunks only to show what overlapping recovers.
A chunk of \(N_y \ell\) rows and \(N_p \ell\) columns gives at least \((N_y - N_p)\ell = 6\ell\) laser free channels, and is equal when \(\bm M\) has full column rank \citep{vallisneri_tdi-infinity_2021, houba_robust_2024}.

This count does not include two costs at each chunk boundary, both of which depend on the interpolator.
A measurement reads laser noise at earlier times, so the columns of each chunk must extend back by a pad of \(d\) samples, which is the largest one-way travel time between spacecraft \(\sim 8.3\,\mathrm{s} \sim 34\; \mathrm{samples}\) at 4\,Hz plus the number of samples the interpolation kernel reaches before the delayed time.
This gives \(d=57\) for the Lagrange with order 45 (46 taps), \(d = 44\) for the cosine-sum with 22 taps, and \(d=34\) for the \ac{GP} whose weights cover the whole chunk and error is absorbed by the covariance rather than padding.
As chunks are factorised independently the \(N_p d\) columns are new unknowns even though the previous chunk contains the same noise, and as their local measurements are in the previous chunk they cannot be cancelled in this one.
Before the first chunk there is no data, so we extend the columns back by \(d\) as well, this costs \(N_p d\) unknowns rather than dropping \(N_y d\) rows.
As well as the column padding, Model 1 (Section~\ref{sec:realistic:model1}) applies the anti-alias filter \(\bm{F}_4\) with \(L\) taps to \(\bm M\), so the first \(N_{\text{trim}} = L - 1 = 53\) rows of each chunk contain its edge transient and are removed.
Each filtered measurement is the sum over the previous \(L\) measurements, the first \(N_{\text{trim}}\) rows rely on measurements that are before the chunk start.
The \ac{GP} kernel of Model 2 (Section~\ref{sec:realistic:model2}) contains the filter and carries the edge error in its covariance instead so \(N_{\text{trim}} = 0\).
Neither \citet{vallisneri_tdi-infinity_2021}
nor \citet{houba_robust_2024} filter \(M\), so there is no row loss there.

Each boundary then costs \(N_y N_{\text{trim}}\) rows and \(N_p d\) columns and the number of laser free channels falls with every trim.
Overlapping the chunks by \(N_{\text{trim}}\) recovers the rows, but the overlapped rows mean that there are also extra columns, so each boundary adds \(N_p(N_{\text{trim}} + d)\) columns.
Fig.~\ref{fig:chunk_trim} shows two chunks of 125 samples of the design matrix for the two models of Section~\ref{sec:realistic}.
In Model 1 the filter is incomplete at the edge, therefore each chunk after the first is extended back by \(N_{\text{trim}}\) rows so that when trimmed the filter is complete at each sample.
In Model 2 there is no overlap as the \ac{GP} includes edge effects in its covariance, so it only pads the columns.

\begin{figure}
    \centering
    \includegraphics[width=\linewidth]{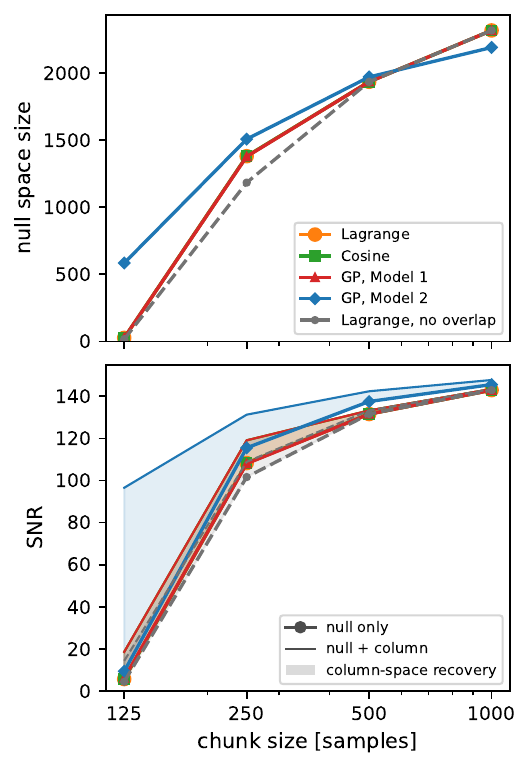}
    \caption{A fixed 1000 sample segment split into
        1, 2, 4 and 8 equal chunks, with all 18 primary noises modelled and the data
        and noise realisation held fixed. The top panel shows the null space size, out of a total of \(24 \times 1000\) measurements.  Bottom, the optimal \ac{SNR} of the injected
        galactic binary, where the markers are the null space only \(\rho_0\) and the lines
        the reweighted \(\rho = (\rho_0^2 + \rho_r^2)^{1/2}\).  The shading between them is what the
        column space recovers.
        The Model 1 configurations are overlapped by \(N_{\text{trim}}\), the grey dashed line is a Lagrange interpolated run without the overlap.
        \ac{SNR} values are for the injection of Table~\ref{tab:noises}.}
    \label{fig:chunk_vs_snr}
\end{figure}

Fig.~\ref{fig:chunk_vs_snr} splits a fixed 1000 sample segment into  1, 2, 4 and 8 chunks, with the same data and noise realisation for each.
The counts are below \(6 \ell\) as the null space is taken after whitening, which keeps only \(N_w\) of the \(N_y \ell\) rows as in Section~\ref{sec:likelihood}.
The lower panel shows the optimal \ac{SNR} of a galactic binary drawn from Table~\ref{tab:noises}, split using equations~(\ref{likelihood:null}) and~(\ref{likelihood:column}) into \(\rho^2 = \rho_0^2 + \rho_r^2\) with
\begin{equation}
    \begin{split}
        \rho_0^2 & = \left\lVert \bm{U}_0^T \bar{\bm{h}} \right\rVert^2 , \\
        \rho_r^2 & = \left(\bm{U}_r^T \bar{\bm{h}}\right)^T \bm{C}_r^{-1}
        \left(\bm{U}_r^T \bar{\bm{h}}\right) ,
    \end{split}
\end{equation}
which are the null and column space contributions.
Both \(\rho\) and \(\rho_0\) fall with chunk number.
The gap between them is the \ac{SNR} recovered by including the column space, which matters most for short chunks and its size depends on the null space threshold of Appendix~\ref{app:threshold}.
At 125 samples Model 2 keeps around 2/3 of its single chunk \ac{SNR}, where the majority comes from column space contributions, whereas the overlapped Model 1 configurations keep significantly less.
The overlap recovers the trimmed rows, but the extra \(N_p N_{\text{trim}}\) columns it adds at each boundary mean signal information gets absorbed into the column space laser prior.
Running Model 2 with the same overlap and trim recovers the same \ac{SNR} as Model 1.

Each chunk is whitened and factorised independently, so the likelihood is a product over chunks and the noise covariance is block diagonal.
Noise correlated on timescales longer than a chunk, i.e.\ at frequencies below \(\sim 1/(\ell\,\Delta t)\), is modelled within each chunk but its correlation between chunks is lost.
A complete analysis would also need to account for these correlations, which we leave to future work.

\section{Parameter estimation tests}
\label{sec:pe}

Sections~\ref{sec:cancel} and~\ref{sec:chunksize} measured the residual laser noise that each
interpolator leaves and the number of laser free channels that it keeps.
Here we measure how they affect the posteriors of the galactic binaries of Section~\ref{sec:data:signal}, injected into the full constellation data of Section~\ref{sec:data:full}.

Each configuration in Table~\ref{tab:pe_cases} changes one setting from the reference, and each uses the same seed for the 200 injections and noise realisations.
Each is analysed with the configurations of Sec.~\ref{sec:realistic} from Sec.~\ref{sec:interpolation}, using the null space likelihood equation~(\ref{likelihood:null}) and the full likelihood of equation~(\ref{likelihood:column}).
The posteriors are sampled using emcee \citep{emcee} using 32 walkers for 4000 steps, with 1000 burn-in steps.
The configuration files and results are available on Zenodo \citep{bayley_tdiinf_zenodo_2026}.

\begin{table}
    \centering
    \caption{Parameter estimation configurations.  \(f_2\) is the stopband edge of the
        anti-alias filter of Section~\ref{sec:data:aafilter}, and all runs use the noises of
        Table~\ref{tab:noises} unless stated. }
    \label{tab:pe_cases}
    \begin{tabular}{lccl}
        \hline
        Name               & Chunks         & \(f_2\) [Hz] & Change from reference \\
        \hline
        Reference          & \(1\times500\) & 1            & ---                   \\
        Stopband 2\,Hz     & \(1\times500\) & 2            & stopband at Nyquist   \\
        Short chunks       & \(2\times250\) & 1            & chunking              \\
        Gaps \(3\times32\) & \(1\times500\) & 1            & 3 gaps of 32 samples  \\
        Gaps \(32\times3\) & \(1\times500\) & 1            & 32 gaps of 3 samples  \\
        \hline
    \end{tabular}
\end{table}

The reference analyses a whole 500 sample segment as a single chunk with no gaps and the stopband at 1\,Hz which is the best case for each interpolator.
As in Section~\ref{sec:cancel:fulldesign} all configurations cancel below the secondary noise at 1\,Hz.
We include the test with the stopband edge at 2\,Hz to test primarily the \ac{GP} in that case, we expect the Lagrange and cosine-sum to fail due to interpolation error in Section~\ref{sec:cancel:interp}.
We also repeat the chunked tests from Section~\ref{sec:chunksize}, looking at data split into two chunks of 250.

Real data will contain gaps, under the assumption that the gaps are in the telemetry samples only and not in the underlying noise, TDI-\(\infty\) handles these by removing rows, while standard TDI combinations lose further samples around each gap as each combination spans several delays \citep{Burke:2025fvl}.
The gaps runs remove rows at the gap samples from \(\bm{y}\), \(\bm{M}\) and the rows and columns of \(\tilde{\bm{C}}_n\) in every channel, while the columns are kept.
There are two cases to test: a few large gaps and many small gaps.

We check the calibration of the posteriors with p-p plots and \ac{KS} tests as in \citet{romero-shaw_bayesian_2020}, where for calibrated posteriors the fraction of posterior samples below the injected value is uniform over the injections.
For each analysis we show the smallest \ac{KS} p-value of \(A, f, \phi_0\) and count it as failing when \(p < 0.001\).
We also measure the size of the posteriors with the posterior volume defined by
\begin{equation}
    \label{eq:pe:volume}
    V = \sqrt{\det \Sigma} ,
\end{equation}
where \(\Sigma\) is the \(3\times3\) sample covariance of the posterior over \((A, f, \phi_0)\).

Each of the five data setups of Table~\ref{tab:pe_cases} is analysed with the four models, each with the null space likelihood and full likelihood with the column space, and each analysis is run on all 200 injections.

\begin{figure*}
    \centering
    \includegraphics[width=\textwidth]{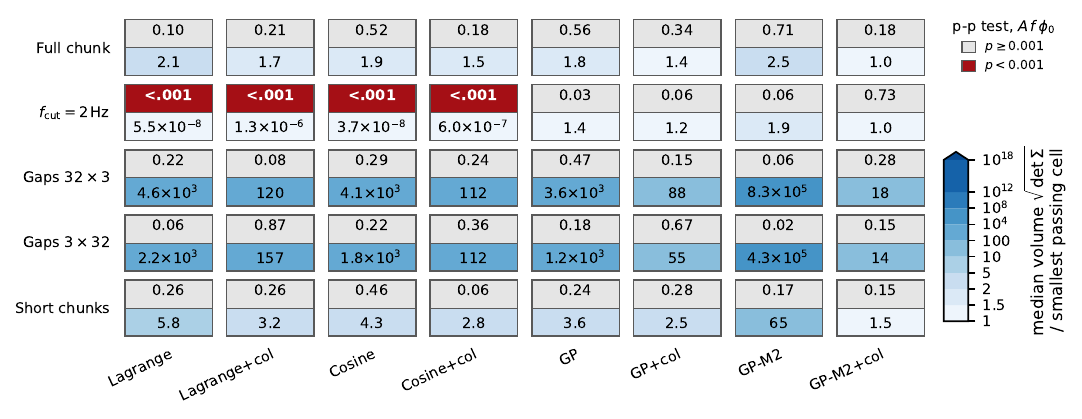}
    \caption{Calibration and posterior size for every parameter estimation run. Rows are the configurations of Table~\ref{tab:pe_cases} and columns the four models, each with the null-space likelihood and with the column-space term added (`+col').
        The upper half of each cell is the p-p test over the 200 injections, where we show the smallest of the \ac{KS} p-values for \(A\), \(f\) and \(\phi_0\).
        Cells with (\(p < 0.001\)) fail and are red. The lower half is the median over injections of the posterior volume \(V = \sqrt{\det\Sigma}\), divided by the smallest median among the passing cells, so a value of 1 marks the smallest calibrated posteriors.}
    \label{fig:pe_overview}
\end{figure*}

Fig.~\ref{fig:pe_overview} summarises all the runs, with one cell per analysis, the data setups in the rows and the interpolator and likelihood in the columns.
The top half of each cell is the smallest of the three \ac{KS} p-values from the p-p tests of \(A\), \(f\) and \(\phi_0\), so a cell passes only when all three parameters are calibrated.
It is shaded red when \(p < 0.001\), which signifies a failed test.
The bottom half is the median of \(V\) over the 200 posteriors, as a ratio to the smallest median of the cells that pass.

In the reference run all eight runs pass the p-p plot \ac{KS} test, i.e.\ \(p \ge 0.001\).
The column space term reduced \(V\) by a factor of 1.3 for Lagrange, cosine-sum and the \ac{GP} with Model 1, and about 2.5 for the \ac{GP} in Model 2.
This gain comes from directions just above the null space threshold, and with a larger threshold the null space posteriors reach the same size (Appendix~\ref{app:threshold}).
The \ac{GP} with Model 2 and the column space term gives the smallest posteriors, although its null space posterior is the largest of the models.

\begin{figure}
    \centering
    \includegraphics[width=1.0\linewidth]{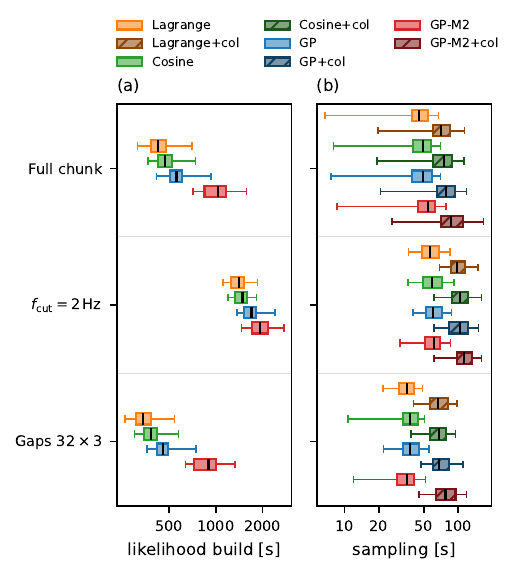}
    \caption{Timings of three of the runs split into the likelihood build and the sampling time for each of the interpolators.}
    \label{fig:timings}
\end{figure}

In the higher stopband test at 2\,Hz, the Lagrange and cosine-sum tests fail the p-p plots, which is expected due to the interpolation error described in Section~\ref{sec:cancel:interp}.
Their posterior volumes are also many orders of magnitude smaller than the reference, so the residual laser noise causes strongly overconfident posteriors.
Both \ac{GP} methods remain calibrated and return results consistent with the full chunk.
The extra bandwidth does not make the \ac{GP} posteriors worse, although the likelihood build does take longer (Fig.~\ref{fig:timings}), and it also adds little in this case as the signal injections are all between 0.02 and 0.05\,Hz.

In the test with two chunks of 250 samples, the posteriors of the Model 1 interpolators are a few times larger than the reference, and the column space recovers part of that.
The \ac{GP} in Model 2 has a larger null space posterior, but with the column space it is within a factor of two of the reference.
We do expect some loss in information as each chunk boundary adds new unknowns, and with Model 1, rows are also trimmed.

In the two gap cases we expect loss of information due to loss of data.
There are two cases: 32 short gaps of size 3 samples and 3 gaps of size 32 samples.
In both cases the null space posteriors of the Model 1 interpolators increase by about 3 orders of magnitude compared to the reference, the column space brings that down to around 2 orders.
Most of this difference also depends on the threshold, however in the case of gapped data, the full posteriors become overconfident before the null space reaches the same volume, so we retain some gain from using the column space (Appendix~\ref{app:threshold}).
The \ac{GP} with Model 2 and the column space term gives the smallest posteriors, which are around an order of magnitude smaller than the Lagrange equivalent in Model 1.
However, its null space posterior alone is significantly larger.


In Fig.~\ref{fig:timings} we show the time it took to build each of the likelihoods, including the design matrix, and the sampling time.
In all cases the likelihood build dominates over the sampling time for these injections.
The \ac{GP} with Model 1 takes slightly longer than Lagrange or cosine-sum to build, with the \ac{GP} in Model 2 taking about 2.5 times as long in the reference run.

With the column space term the \ac{GP} with Model 2 gives the smallest posteriors in every run and the \ac{GP}, with either model, is the only interpolator that stays valid when including information up to the Nyquist of 2\,Hz.
The gain of the \ac{GP} and the column space is small in the runs of a single large chunk when the laser noise is much larger than the secondary noise, as is expected for \ac{LISA}.
However when the data is gapped, the \ac{GP} can achieve an order of magnitude smaller posteriors when including the column space.
How much the column space adds depends on the choice of null space threshold, for a single chunk a larger threshold recovers the same posteriors with the null space alone, whereas with gaps part of the gain remains (Appendix~\ref{app:threshold}).

\section{Conclusion}

In this work we extend TDI-\(\infty\) from \citet{houba_robust_2024} to data that has passed through the on-board anti-alias filter and has been downsampled to the telemetry rate.
The design matrix for this cannot be built at the very high rate the laser noise is sampled, so we compare two effective models at the telemetry rate (Section~\ref{sec:realistic}).
In Model 1, used with all three interpolators, the filter is applied after the delays, following the order on the spacecraft but we remove the first \(L-1\) rows from every chunk.
In Model 2, with just the \ac{GP}, the unknowns are the filtered noise samples, and the \ac{GP} interpolates this filtered noise directly to the delays times.
The filter is built into the kernel and the interpolation error is added to the secondary noise covariance through \(C_\epsilon\).
Alongside this we also compare (i) using just the null space likelihood as in previous implementations, with (ii)keeping the column space components in Section~\ref{sec:likelihood}.

Of the three interpolators, the \ac{GP} reproduces the filtered noise most accurately between 1 and 2\,Hz, as its kernel is built from the filter that produced the data (Section~\ref{sec:cancel}).
The Lagrange and cosine-sum errors are largest between 1 and 2\,Hz, and when the stopband edge of the filter is moved above 1\,Hz the residual laser noise rises by several orders of magnitude.
On the full data simulation, all three interpolators cancel the laser noise below the secondary noise when the stopband is set to 1\,Hz, but only the \ac{GP}, in either model, does so when the stop band is at 2\,Hz.

Splitting the design matrix into chunks costs a column pad at every boundary for all interpolators and in Model 1 the first \(N_{\text{trim}}\) rows of the filter transient are dropped (Section~\ref{sec:chunksize}).
At 125 samples per chunk Model 1 has almost no null space left, and overlapping chunks recovers the samples but not the laser free channels.
In contrast, the \ac{GP} in Model 2 maintains much of its null space, and the column space increases the \ac{SNR} by a factor of 10 at our null space threshold.

In the parameter estimation tests in Section~\ref{sec:pe}, we analysed 200 `galactic binary'-like injections in five data setups.
In the reference setup we use a single 500 sample chunk with a 1\,Hz stop band edge, all models give calibrated posteriors for this setup.
The column space reduces the posterior volume by a factor of around 1.3 in Model 1 and 2.5 in Model 2, and the \ac{GP} in Model 2 with the column space gives the smallest posteriors, although a larger null space threshold would recover most of this gain with the null space alone.
With two chunks of 250 samples, the Model 1 posteriors are a few times larger than the reference, while the \ac{GP} in Model 2 with the column space stays within a factor of two.
All models stay calibrated with gaps, which TDI-\(\infty\) handles by removing rows.
The Model 1 null space posteriors grew by three orders of magnitude, and including the column space reduces this to grow by two orders, some of which remains for any null space threshold.
When using the \ac{GP} in Model 2 with the column space, the posteriors are an order of magnitude smaller than the Lagrange and cosine-sum posteriors, and several times smaller than the \ac{GP} in Model 1.
This loss is far larger than one might expect from the proportion of samples removed, because each gap also prevents cancellation in the measurements within roughly a light travel time of its edges. 
For short segments with closely spaced gaps that affects most of the data and the impact should be much smaller for well-separated gaps and longer segments.
The size of the null space loss also depends on the threshold (App.~\ref{app:threshold}), whereas the full posteriors only change when the threshold is raised enough to cause the posteriors to collapse. 
The \ac{GP} takes \(\sim 1.3\) times longer than the Lagrange in Model 1 to build the likelihood, and the Model 2 \ac{GP} takes about \(\sim 2.5\) times longer.

These tests were run on a simple astrophysical model with unrealistically high amplitudes and short data segments.
In the future a test with more realistic astrophysical signals could be instructive.
Also further testing on computing the \ac{GP} design matrix using double precision would be needed as using the emulated float128 of Section~\ref{sec:realistic:model2} is not practical for full \ac{LISA} data.
Further to this, for a full analysis one would have to consider the non-zero correlations between chunks.
The null space threshold could also be set from the primary noise prior rather than as a fraction of the largest singular value.

In summary, we have shown that current implementations of TDI-\(\infty\) can be extended to filtered and downsampled telemetry data provided the on-board filter is built into the design matrix.  We recommend that a \ac{LISA} analysis based on TDI-\(\infty\) use an interpolator that remains accurate up to the Nyquist frequency and accounts for its own error, as the \ac{GP} does, and keep the column space of the design matrix whenever the data are split into chunks or contain gaps.

\section*{Acknowledgements}
We would like to thank Jean-Baptiste Bayle and Niklas Houba for discussions on the realistic data model and TDI-\(\infty\), as well as help supplying initial code on a simplified model.
We would also like to thank Christopher Messenger, Christopher Berry and John Veitch for various discussions throughout this work.
Thanks to Daniel Williams for a recommendation on the title.
This work was supported by grant UKRI972 awarded via the UK
Space Agency.

\section*{Data Availability}
Code and configuration file are available on Zenodo \citep{bayley_tdiinf_zenodo_2026}.

\bibliographystyle{mnras}
\bibliography{refs}

\appendix
\section{Choice of the secondary noise cut}
\label{app:cncut}

\begin{figure}
    \centering
    \includegraphics[width=\linewidth]{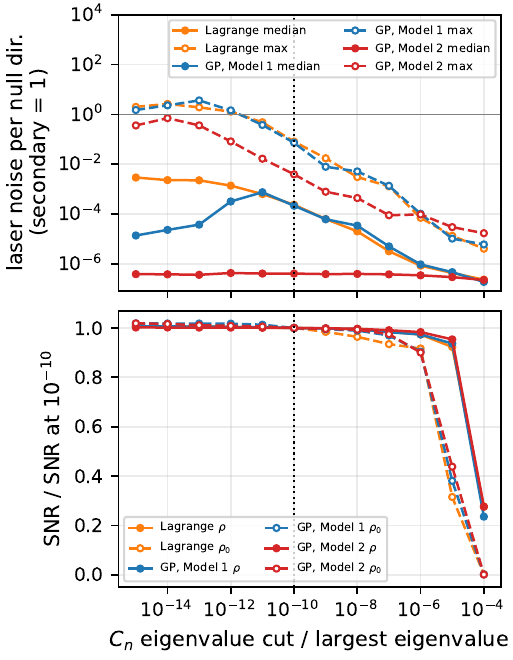}
    \caption{\label{fig:cncut} Choice of the relative cut on the eigenvalues of
        \(\tilde C_n\), for one 500 sample chunk with a \(1\,\mathrm{Hz}\) stopband edge.  Top,
        residual laser noise per laser-free channel in units of the secondary noise, median
        (solid) and maximum (dashed) over channels.  Bottom, the \ac{SNR} of a galactic binary
        for the full likelihood \(\rho\) (solid) and the null space alone \(\rho_0\) (dashed),
        relative to their values at the \(10^{-10}\) cut (dotted line).}
\end{figure}

The whitening to the secondary noises as in Section~\ref{sec:likelihood} keeps only the eigenvectors of \(\tilde{\bm{C}_n}\) whose eigenvalues are above a fraction of the largest eigenvalue.
We show in Fig.~\ref{fig:cncut} how changing this fraction affects the residual laser noise and the \ac{SNR} of a signal.
We set the cut to \(10^{-10}\), the lowest value where the largest residual laser noise is a order of magnitude lower than the secondary noise for every model. the largest residual is just below the secondary noise. 
Higher cuts would also be acceptable, i.e. at \(10^{-8}\) the largers residual is an order of magnitude lower than at \(10^{-10}\) with only a small loss in \ac{SNR}.

\section{Choice of singular value threshold cut}
\label{app:threshold}

In the above sections we define the null space with a threshold of \(10^{-14}\) of the largest singular value of \(\bar{\bm{M}}\), which is the round off floor of the \ac{SVD}.
Above this floor there are directions where the laser noise is cancelled well below the secondary noises but not exactly, which can be seen in the lower panel of Fig.~\ref{fig:singular_tdiinf}.
The null space likelihood discards those directions, whereas the column space keeps them weighted by the primary noise prior.
The null space posteriors and the gain from the column space depend on the choice of threshold.

\begin{figure}
    \centering
    \includegraphics[width=1.0\linewidth]{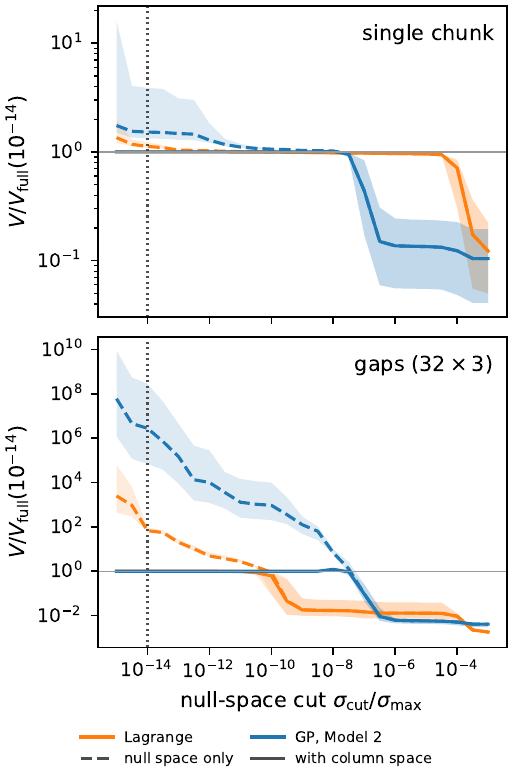}
    \caption{Posterior volume against the relative null space threshold, for one 500 sample chunk (top) and 32 gaps of size 3 samples (bottom). Both the Lagrange (Model 1) and \ac{GP} (Model 2) use the 1\,Hz stopband edge. Each injection's volume is divided by its full likelihood volume at the \(10^{-14}\) threshold (vertical dotted line), for the null space likelihood (dashed) and the null+column (solid). Lines are the medians over the injections and bands are the 1\(\sigma\) spread.}
    \label{fig:null_threshold}
\end{figure}

In Fig.~\ref{fig:null_threshold} we show how the posterior volume of equation ~\ref{eq:pe:volume} changes with the threshold for a single chunk and for 32 gaps of size 3.
For a single chunk, raising the threshold moves these directions into the null space and the null space posterior shrinks to the size of the full posterior.
Only at thresholds above \(10^{-7}\) does the \ac{GP} model begin to give smaller over constrained posteriors due to un-cancelled laser noise leaking into the null space.
With gaps the full posterior begins to collapse before the null space posterior reaches it, so part of the column space information cannot be recovered at any threshold.
Where this happens depends on which model we are using, so the threshold would have to be chosen depending on which pipeline was used.

\bsp
\label{lastpage}
\end{document}